\documentclass[prl,twocolumn,groupedaddress,noshowpacs,nofootinbib]{revtex4}
\usepackage{graphicx}
\usepackage{epsfig}

\def\nd{$\nu\phantom{}DE$}
\def\aa{{\cal A}}
\begin{document}
\title{Left-Right Symmetric Model of Neutrino Dark Energy}

\author{Jitesh R. Bhatt$^{1}$}
\email{jeet@prl.res.in}

\author{Pei-Hong Gu$^{2}_{}$}
\email{pgu@ictp.it}

\author{Utpal Sarkar$^{1}_{}$}
\email{utpal@prl.res.in}

\author{Santosh K. Singh$^{1}_{}$}
\email{santosh@prl.res.in}

\affiliation{$^{1}_{}$Physical Research Laboratory, Ahmedabad 380009, India\\
$^{2}_{}$The Abdus Salam International Centre for
Theoretical Physics, Strada Costiera 11, 34014 Trieste, Italy}

\begin{abstract}\

We implemented the neutrino dark energy (\nd) proposal in a
left-right symmetric model. Unlike earlier models of mass varying
neutrinos, in the present model the mass parameter that depends on
the scalar field (acceleron) remains very light \emph{naturally}.
The required neutrino masses then predicts the $U(1)_R$ breaking
scale to be in the TeV range, providing new signals for LHC.
Compared to all other \nd~proposals, this model has the added
advantage that it can also be embedded into a grand unified theory.
In this scenario leptogenesis occurs through decays of scalars at
very high energy.
\end{abstract}

\maketitle

Present observations reveal that the dark energy $\sim(3\times
10^{-3}\,\textrm{eV})^{4}$\cite{dark} contributes about $70\%$ to
the total density of our universe. Since the only known physics
around this scale is the neutrino mass, there are now attempts to
relate the origin of the dark energy with the neutrino masses
\cite{mavans1,mavans2,mavans3}. This connection is based on the idea
of quintessence \cite{quint}, and have several interesting
consequences \cite{kap,pas}.

In the original model of neutrino dark energy (\nd) or the mass
varying neutrinos (mavans) \cite{mavans1,mavans2,mavans3}, the
standard model is extended by including singlet right-handed
neutrinos $N_i, i=1,2,3$, and giving a Majorana mass to the
neutrinos which varies with a scalar field, the acceleron. This
model was not complete and several problems were pointed out
\cite{mavans2,azk2005}. Some of the problems have been solved in
subsequent works \cite{models,tripnd}, but more studies are required
to make this model fully consistent. The main motivation of the
present article is to justify the very low scale entering in this
model naturally, embed this idea into a left-right symmetric model
and also in grand unified theories. Since the right-handed neutrinos
are not very heavy, leptogenesis occurs through scalar decays.

In the \nd~models, the Majorana masses of the right-handed neutrinos
varies with the acceleron field and that relates the scale of dark
energy with the light neutrino masses. Naturalness requires the
Majorana masses of the right-handed neutrinos also to be in the
range of eV, so the main motivation of the seesaw mechanism is lost.
The smallness of the light neutrino masses cannot be attributed to a
large lepton number violating mass scale in the theory. In this
\nd~model, the neutrino Dirac masses cannot be made to vary with the
acceleron field, since that will then allow coupling of the
acceleron field with the charged leptons and a natural scale for the
dark energy will then be the mass of the heaviest charged lepton.
For the same reason, this mechanism cannot be embedded into a
left-right symmetric model, in which the $SU(2)_R$ group relates the
right-handed neutrinos to the right-handed charged leptons.

The problem with the smallness of the mass parameter that depends on
the acceleron field can be softened in the \nd~models with triplet
Higgs scalars \cite{tripnd}. In these models the standard model is
extended to include triplet Higgs scalars. In any phenomenologically
consistent triplet Higgs scalar model, lepton number is violated
explicitly by a trilinear scalar couplings of the triplet Higgs
scalar with the standard model Higgs doublet. In the \nd~model with
the triplet Higgs scalars, the coefficient of this trilinear scalar
coupling with mass dimension varies with the acceleron field, and
naturalness allows this parameter to be as large as a few hundred
GeV. Although the scale of this mass parameter predicts new signals
in the TeV range, there is no symmetry that makes this scale
natural.

We propose a left-right symmetric model, in which the mass parameter
that varies with the acceleron field remains small naturally and the
scale of dark energy is related to the neutrino masses. This is the
only \nd~model that can be embedded into a grand unified theory,
without relating the scale of dark energy to the charged fermion
masses. We then discuss the question of leptogenesis in this model.

We start with the left-right symmetric extension of the standard
model \cite{lr} with the gauge group $G_{LR} \equiv SU(3)_c \times
SU(2)_L \times SU(2)_R \times U(1)_{B-L}$, and the electric charge
is related to the generators of the group as:
\begin{equation}\label{1}
    Q = T_{3L} + T_{3R} + {B-L \over 2} = T_{3L} + Y\,.
\end{equation}
The quarks and leptons transform under the left-right symmetric
group as:
\begin{eqnarray}
Q_L =\pmatrix{ u_L \cr d_L }  \equiv [ 3,2,1, {1 \over 3} ]&& Q_R
=\pmatrix{ u_R \cr d_R }  \equiv [ 3,1,2, {1 \over 3}]
\nonumber \\
\ell_L  =\pmatrix{ \nu_L \cr e_L }  \equiv [ 1,2,1, {-1  } ]&&
\ell_R  =\pmatrix{ N_R \cr e_R }  \equiv [ 1,1,2, -1]
\nonumber \\
S_R \equiv [1,1,1,0]\,.
\end{eqnarray}
In addition to the standard model fermions, the right-handed
neutrinos $N_R$ and a right-handed singlet fermion
$S_R$ have been introduced. Under left-right parity this
field transform to its $CP$ conjugate state
as: $S_R \leftrightarrow {S^c}_L$, and the
Majorana mass term is invariant under the parity transformation.
So, although we do not include another field $S_L$, the theory
is left-right symmetric. This is possible because this field
transform to itself $S_R \equiv (1,1,1,0) \leftrightarrow
(1,1,1,0)$ under the transformation $SU(2)_L
\leftrightarrow SU(2)_R$.

We consider the symmetry breaking pattern \cite{rajpoot1981,v}:
\begin{eqnarray}
&&SU(3)_c \times SU(2)_L \times SU(2)_R \times U(1)_{(B-L)} ~[
G_{LR}] \nonumber \\ & \stackrel{M_R}{\rightarrow} &SU(3)_c \times
SU(2)_L \times U(1)_R \times U(1)_{(B-L)} ~~~[ G_{1R}] \nonumber \\
& \stackrel{m_r}{\rightarrow} &SU(3)_c \times
SU(2)_L \times U(1)_Y ~~~~~~~~~~~~~~~~~~~~\,[ G_{std}] \nonumber \\
&\stackrel{m_W}{\rightarrow} &SU(3)_c \times U(1)_Q ~~~
~~~~~~~~~~~~~~~~~~~~~~~~~~~~~~~[ G_{em}]\,.  \nonumber
\end{eqnarray}
The Higgs scalars required to break the left-right symmetric
group to $G_{1R}$ transform as $\xi_R \equiv (1,1,3,0)$.
This Higgs scalar does not couple to the fermions and cannot
give Majorana masses to the neutrinos, since it does not carry
any $B-L$ quantum number. The
group $G_{1R}$ and the $B_L$ symmetry
is broken by the vacuum expectation value ($vev$) of
the field $\chi_R \equiv (1,1,2,1)$ \cite{doub1,albright}.
If $SU(2)_R$ is not broken to its subgroup $U(1)_R$ at some
high scale, the field $\chi_R$ can break the left-right
symmetry group directly to the standard model.
For consistency with the left-right
symmetry, or the existence of the left-right parity
would then require the fields $\xi_L \equiv (1,3,1,0)$
and $\chi_L \equiv (1,2,1,1)$. We break the standard model
gauge symmetry by a bi-doublet $\Phi \equiv (1,2,2,0)$, whose
$vev$ can give masses to the charged fermions. In addition, we
introduce another bi-doublet $\Psi \equiv (1,2,2,0)$, which
does not contribute to the fermion masses but has similar $vev$ and
a singlet scalar field $\eta \equiv (1,1,1,0)$, which acquires a tiny
$vev$ and generate the mass scale for the dark energy.

We start with the interactions of the Higgs scalar fields. There
are quadratic and quartic self interactions of all the fields,
which determines their masses and vacuum expectation values ($vev$).
However, some of the fields would acquire induced $vev$s due to their
linear interactions. We shall first write down these terms which will
allow us to determine the $vev$s of the different fields. In principle,
one should write down all the scalar interactions and then minimize
the potential to find the consistent solution for the $vev$s of the
different fields. These details will be presented elsewhere.
Here we shall present the essential part of the scalar interactions and
an estimate of the $vev$s. In addition to the usual
quadratic and quartic interactions
of the different fields, for the working of the
present mechanism the Lagrangian contains the terms:
\begin{eqnarray}
  {\cal L}_s &=& h_\Phi \eta (\Psi \xi_L \Phi
  + \Psi^\dagger \xi_R \Phi)
  + h_\chi \eta \chi_L^\dagger \chi_R \Phi \nonumber \\
  & +& h_\xi \xi_L \xi_R (\Phi^\dagger \Phi + \Psi^\dagger \Psi)
\label{scalar}
\end{eqnarray}
This Lagrangian results from a $Z_4$ discrete
symmetry, under which the different fields transform as:
$$ \begin{array}{c@{\hspace{0.5in}}c@{\hspace{0.5in}}c}
\chi_L \to i \chi_L & \chi_R \to -i \chi_R & S_R \to i S_R \\
\xi_L \to i \xi_L & \xi_R \to -i \xi_R & \eta \to - \eta \\
& \Psi \to i \Psi&
\end{array} $$
Denoting the $vev$s of the different fields by:
\begin{eqnarray}
  \langle \xi_L \rangle = u_L && \langle \xi_R \rangle = u_R \nonumber \\
  \langle \chi_L \rangle = v_L && \langle \chi_R \rangle = v_R \nonumber \\
  \langle \Phi \rangle = v && \langle \eta \rangle = u  \nonumber \\
  \langle \Psi \rangle = w && \nonumber
\end{eqnarray}
we can minimize the complete scalar potential and find a consistent
solution with (the details will be presented elsewhere):
\begin{eqnarray}
   \begin{array}{c@{\hspace{0.1in}}c@{\hspace{0.1in}}c}
        u \approx \displaystyle{v w (u_L +u_R) \over m_\eta^2} & u_L \approx
        \displaystyle{(v^2 + w^2) u_R \over m_\xi^2} & v_L \approx
        \displaystyle{v v_R u \over m_\chi^2}
    \end{array}&& \nonumber \\
      u_R \gg v_R > v > w \gg u \gg v_L\,.\phantom{push hard}
\end{eqnarray}
In grand unified theories the consistency of the gauge coupling
unification requires the scale of left-right symmetry breaking to be
above $10^{11}$~GeV, so we shall assume $u_R \sim 10^{11}$~GeV. We
also assume $m_\eta \sim m_\xi \sim u_R$. However, the $G_{1R}$
symmetry breaking scale could be very low, so we shall assume
$m_\chi \sim v_R \sim $~TeV. The other mass scales are then $v \sim
m_w \sim 100$~GeV, $u \sim u_L \sim $~eV and $v_L \sim 10^{-2}$~eV.
Since the $B-L$ symmetry is broken around the TeV scale, there will
be new phenomenological consequences that may be observed at LHC.

The neutrino masses come from the
Yukawa interactions of the leptons and the singlet fermion $S$,
which are given by:
\begin{eqnarray}
  {\cal L}_Y &=& f \overline{\ell_L}~ \ell_R \Phi + f_L \overline{S_R} ~\ell_L \chi_L
                 + f_R \overline{{S^c}_L} \ell_R ~\chi_R \nonumber \\
   &+& \frac{1}{2}f_s \eta \overline{{S^c}_L}~ S_R + H.c.\,.
\label{yuk}
\end{eqnarray}
The Yukawa couplings $f$ are $3 \times 3$ matrix, while $f_L$ and
$f_R$ are $3 \times n$ matrices, if we assume that there are $n$
singlet fermions $S$ and $f_s$ is a $n \times n$ matrix. The
neutrino mass matrix can now be written in the basis
$\pmatrix{\nu_{L} & {N^c}_L & {S^c}_L}$ as:
\begin{equation}
M_\nu = \pmatrix { 0 & f v & f_{L} v_L \cr f v & 0 & f_{R} v_R \cr
f_{L} v_L & f_{R} v_R & f_s u }\,.
\end{equation}
This matrix can be block diagonalised, which gives the masses of the
right-handed neutrinos and the singlet fermions $S$ to be of the order
of the largest entry in the mass matrix $v_R$. The left-handed
neutrinos remains light with small admixture with heavier states
and the light eigenvalue comes out to be
\begin{equation}
m_{\nu } = - 2{f f_{ L}  \over f_{R}}{ v v_L \over v_R } + {f_s f^2
\over f^2_{R} } {u v^2 \over v_R^2}\,.
\end{equation}
The first term is the type-III seesaw \cite{flhj1988} contribution
and the second term is the double seesaw contribution. With the
choice of the $vev$s discussed earlier, both these terms become
comparable, although the second term dominates.

We now assume that the mass parameter $M_s = f_s u \sim f_s \langle
\eta \rangle$ varies with the acceleron field $\aa$. This parameter
$M_s$ remains of the order of eV \emph{naturally}, and it does not
couple to the charged fermions. Thus the model satisfies both the
conditions we wanted to achieve. Embedding this model in a grand
unified theory is also straightforward. Consider an $SO(10)$ grand
unified theory. The quarks and leptons in this model would belong to
a 16-dimensional representation, while $S_R$ will belong to a
singlet representation. So, the Majorana mass $M_s$ of the singlet
can vary with the acceleron field without affecting the charged
fermion masses. The scalars belong to representations:
$\xi_{L,R}~[45]$, $\chi_L~[16]$, $\chi_R~[\overline{16}]$,
$\eta~[1]$ and $\Phi~[10]$. These fields will then allow the
interactions required for the implementation of this model. When
this model is embedded in a grand unified theory, the different mass
scales for the left-right symmetry breaking and the $U(1)_R$
symmetry breaking come out to be consistent with the gauge coupling
unification.

We shall now discuss the implementation of the \nd~mechanism in this
model. We assume that the singlet mass $M_s$ varies with the
acceleron field $\aa$, so that the neutrino mass becomes a dynamical
quantity since the double seesaw contribution dominates over the
type-III seesaw. This gives the coupling between the neutrinos
and the acceleron, which stops the dynamical evolution of the
acceleron fields when the neutrinos become non-relativistic.
The dependence of the mass $M_s$ on the acceleron field governs
the dynamics of the dark energy. This details would depend on the
nature of the acceleron field. Since we shall not be specifying
the origin of the acceleron field, we shall comment only on
some generic structures of this solution.

As in the original \nd~model, we consider the nonrelativistic limit,
when $m_\nu$ is a function of dark energy, the potential of dark
energy becomes
\begin{equation}
\label{effective-potential} V = m_\nu(\aa)\, n_\nu + V_0 (\aa)\,.
\end{equation}
Here the scalar potential $V_0 (\aa)$ is due to the acceleron field,
for example \cite{mavans2},
\begin{eqnarray}
V_0\,(\aa) &=&  \Lambda^4~ \log (1 + |M_s(\aa)/ \mu |)\,.
\end{eqnarray}
Due to the back reaction from the neutrinos, the evolution of
acceleron field should be described by the effective potential
(\ref{effective-potential}) which depends on the total numbers
$n_\nu$ of thermal background neutrinos and antineutrinos.

The acceleron field will be trapped at the minima of the potential,
which ensures that as the neutrino mass varies, the value of the
acceleron field will track the varying neutrino mass. One generic
feature of this solution is that it leads to a equation of state
with $\omega = -1$ at present. The most important feature of this
scenario is that the energy scale for the dark energy gets
related to the neutrino mass, which is highly desirable. This also
explains why the universe enters an accelerating phase now \cite{wet1}.

The effective low-energy Lagrangian will now become
\begin{equation}
- {\cal L}_{eff} = M_s(\aa) {f^2 \over f_R^2}{v^2 \over v_R^2} ~
\nu_i \nu_j + H.c. + V_0(\aa)\,,
\end{equation}
where $M_s$ is naturally of the order of fraction of eV and hence
can explain the dark energy with the equation of state satisfying $w
=-1$. The scale of dark energy $\Lambda \sim 10^{-3}$ eV does not
require any unnaturally small Yukawa couplings or symmetry breaking
scale in this case. The electroweak symmetry breaking scale $v$ and
the $U(1)_R$ breaking scales are comparable and hence the new gauge
boson corresponding to the group $U(1)_R$ will have usual mixing
with $Z$ and should be accessible at LHC.

Since the minima of the potential relates the neutrino mass to a
derivative of the acceleron potential, the value of the acceleron
field gets related to the neutrino mass. On the other hand, as the
neutrino mass grows, the degeneracy pressure due to the background
neutrinos and antineutrinos also starts growing. This causes problem
with the stability of this solution \cite{azk2005,stab}. However,
this generic problem of this scenario may be explained by
considering formation of neutrino lumps in the universe. As the
neutrino mass grows, there would be a tendency for the neutrinos to
cluster together due to the attractive force originating from the
acceleron coupling. These neutrino lumps would then behave as dark
matter and will not affect the dynamics of the acceleron field,
making the solution stable \cite{wet2}.

In this scenario leptogenesis \cite{fy1986} occurs through decays
of the heavy scalars $\eta$. Unlike other models of type III
seesaw mechanism \cite{albright}, the right-handed neutrinos and
the singlet fermions $S_R$ have masses in the TeV range, and
hence, their decays cannot generate any lepton asymmetry. When
$\xi_R$ acquires its $vev$ at very large scale, the heavy scalars
$\eta$ can decay into $\eta \to \Phi^\ast + \Psi$ and $\eta \to
S_R + S_R$. These decays of $\eta$ can generate a asymmetry in
$S_R$ and ${S^c}_L$ when the tree-level diagrams interfere with
self-energy type one loop diagrams \cite{pasus}. Since $S_R$ does
not carry any $B-L$ quantum number, a lepton asymmetry is not
generated at this time. Before the electroweak symmetry breaking,
when the field $\chi_R$ acquires $vev$, the singlet fermions $S_R$
mix with the right-handed neutrinos and at this time the asymmetry
in $S_R$ and $S^c_L$ is converted to a lepton asymmetry of the
universe. This lepton asymmetry, in turn, generates the required
baryon asymmetry of the universe in the presence of the sphalerons
\cite{krs1985}.

In conclusion, we proposed a left-right symmetric extension of the
standard model, where the \nd~mechanism could be embedded. The most
important advantage of this model over all the existing models is
that it allows a naturally small scale for the dark energy. The
existence of the large scale that generates this small scale
naturally through a seesaw suppression, allows leptogenesis in this
model. The model has the added feature that it can be embedded in a
grand unified theory.


\begin{thebibliography}{99}

\bibitem{dark} A. Riess {\it et al.}, Astron. J. {\bf 116}, 1009 (1998);
S. Perlmutter {\it et al.}, Astrophys. J. {\bf 517}, 565 (1999);
D. Spergel {\it et al.}, Astrophys. J. Suppl. {\bf 148}, 175 (2003).

\bibitem{mavans1} P. Gu, X. Wang, and X. Zhang,
Phys. Rev. D {\bf 68}, 087301 (2003).

\bibitem{mavans2} R. Fardon, A.E. Nelson, and N.
Weiner, JCAP {\bf 0410}, 005 (2004).

\bibitem{mavans3} P.Q. Hung, hep-ph/0010126.

\bibitem{quint} C. Wetterich, Nucl. Phys. B {\bf 302}, 668 (1988); P.J.E.
Peebles and B. Ratra, Astrophys. J. {\bf 325}, L17 (1988).

\bibitem{kap}
D.B. Kaplan, A.E. Nelson, and N. Weiner, Phys. Rev. Lett. {\bf 93},
091801 (2004); H. Li, Z. Dai, and X. Zhang, Phys. Rev. D {\bf 71},
113003 (2005); V. Barger, P. Huber, and D. Marfatia Phys. Rev. Lett.
{\bf 95}, 211802 (2005); A.W. Brookfield, C. van de Bruck, D.F.
Mota, and D. Tocchini-Valentini, Phys. Rev. Lett. {\bf 96}, 061301
(2006); A. Ringwald and L. Schrempp, JCAP {\bf 0610}, 012 (2006).




\bibitem{pas}
R. Barbieri, L.J. Hall, S.J. Oliver, and A. Strumia, Phys. Lett. B
{\bf 625}, 189 (2005); C.T. Hill, I. Mocioiu, E.A. Paschos, and U.
Sarkar, Phys. Lett. B {\bf 651}, 188 (2007); P.H. Gu, H.J. He, and
U. Sarkar, Phys. Lett. B {\bf 653}, 419 (2007); JCAP {\bf 0711}, 016
(2007); P.H. Gu, Phys. Lett. B {\bf 661}, 290 (2008).

\bibitem{azk2005}
N. Afshordi, M. Zaldarriaga, and K. Kohri, Phys.Rev. D {\bf 72},
065024 (2005).


\bibitem{models}
R. Takahashi and M. Tanimoto, Phys. Lett. B {\bf 633}, 675 (2006);
R. Fardon, A.E. Nelson, and N. Weiner, JHEP {\bf 0603}, 042 (2006);
O.E. Bjaelde, {\it et al.}, arXiv:0705.2018 [astro-ph]; N.
Brouzakis, N. Tetradis, and C. Wetterich, arXiv:0711.2226
[astro-ph].


\bibitem{tripnd} E. Ma and U. Sarkar, Phys. Lett. B {\bf 638}, 356 (2006).


\bibitem{lr} J.C. Pati and A. Salam,
Phys. Rev. Lett. {\bf 31}, 661 (1973); Phys. Rev. D {\bf 8}, 1240
(1973); {\it ibid.} {\bf 10}, 275 (1974); R.N. Mohapatra and J.C.
Pati, Phys. Rev. D {\bf 11}, 566, 2558 (1975); R.N. Mohapatra and G.
Senjanovic, Phys. Rev. D {\bf 12}, 1502 (1975).

\bibitem{rajpoot1981}
S. Rajpoot, Phys. Lett. B {\bf 108}, 303 (1981); {\it ibid.} {\bf
115}, 396 (1982).



\bibitem{v} M. Malinsky, J.C. Romao, and J.W.F. Valle, Phys. Rev.
Lett. {\bf 95}, 161801 (2005); M. Hirsch, J.W.F. Valle, M. Malinsky,
J.C. Romao and U. Sarkar, Phys. Rev. D {\bf 75}, 011701 (2007).

\bibitem{doub1} S.M. Barr, Phys. Rev. Lett. {\bf 92}, 101601 (2004).

\bibitem{albright} C.H. Albright and S.M. Barr, Phys. Rev.
D {\bf 69}, 073010 (2004); Phys. Rev. D {\bf 70}, 033013 (2004).


\bibitem{flhj1988}
R. Foot, H. Lew, X.G. He, and G.C. Joshi, Z. Phys. C {\bf 44}, 441
(1989); E. Ma, Phys. Rev. Lett. {\bf 81}, 1171 (1998).


\bibitem{wet1} C. Wetterich, Phys. Lett. B {\bf 655}, 201 (2007).


\bibitem{stab} O.E. Bjaelde, {\it et. al.}, arXiv:0705.2018
[astro-ph].


\bibitem{wet2} D.F. Mota, V. Pettorino, G. Robbers and C. Wetterich,
arXiv:0802.1515 [astro-ph].


\bibitem{fy1986}
M. Fukugita and T. Yanagida, Phys. Lett. B {\bf 174}, 45 (1986); P.
Langacker, R.D. Peccei, and T. Yanagida, Mod. Phys. Lett. A {\bf 1},
541 (1986); M.A. Luty, Phys. Rev. D {\bf 45}, 455 (1992); R.N.
Mohapatra and X. Zhang, Phys. Rev. D {\bf 46}, 5331 (1992).

\bibitem{pasus} M. Flanz, E.A. Paschos and U. Sarkar,
Phys. Lett. B {\bf 345}, 248 (1995); E. Ma and U. Sarkar, Phys. Rev.
Lett. {\bf 80}, 5716 (1998).

\bibitem{krs1985}
V.A. Kuzmin, V.A. Rubakov, and M.E. Shaposhnikov, Phys. Lett. B {\bf
155}, 36 (1985); R.N. Mohapatra and X. Zhang, Phys. Rev. D {\bf 45},
2699 (1992).


\end{thebibliography}
\end{document}